\documentclass[journal=jctcce,manuscript=article,layout=onecolumn]{achemso}

\usepackage[version=3]{mhchem} 
\usepackage{hyperref}
\usepackage{tikz}
\usepackage{subcaption}
\usepackage{graphicx}
\usepackage{tabularray}

\usetikzlibrary{positioning, shapes.geometric, arrows.meta}

\usepackage{soul}

\author{Guangming Liu}
\altaffiliation{These authors contributed equally to this work}
\author{Siwei Wang}
\altaffiliation{These authors contributed equally to this work}
\author{Hsing-Ta Chen}
\email{hchen25@nd.edu}
\affiliation[University of Notre Dame]
{Department of Chemistry\&Biochemistry, University of Notre Dame, Notre Dame, IN, 46616}

\title{\textsc{MQED-QD}: An Open-Source Package for Quantum Dynamics Simulation in Complex Dielectric Environments}

\begin{document}

\begin{tocentry}





\includegraphics[width=\linewidth]{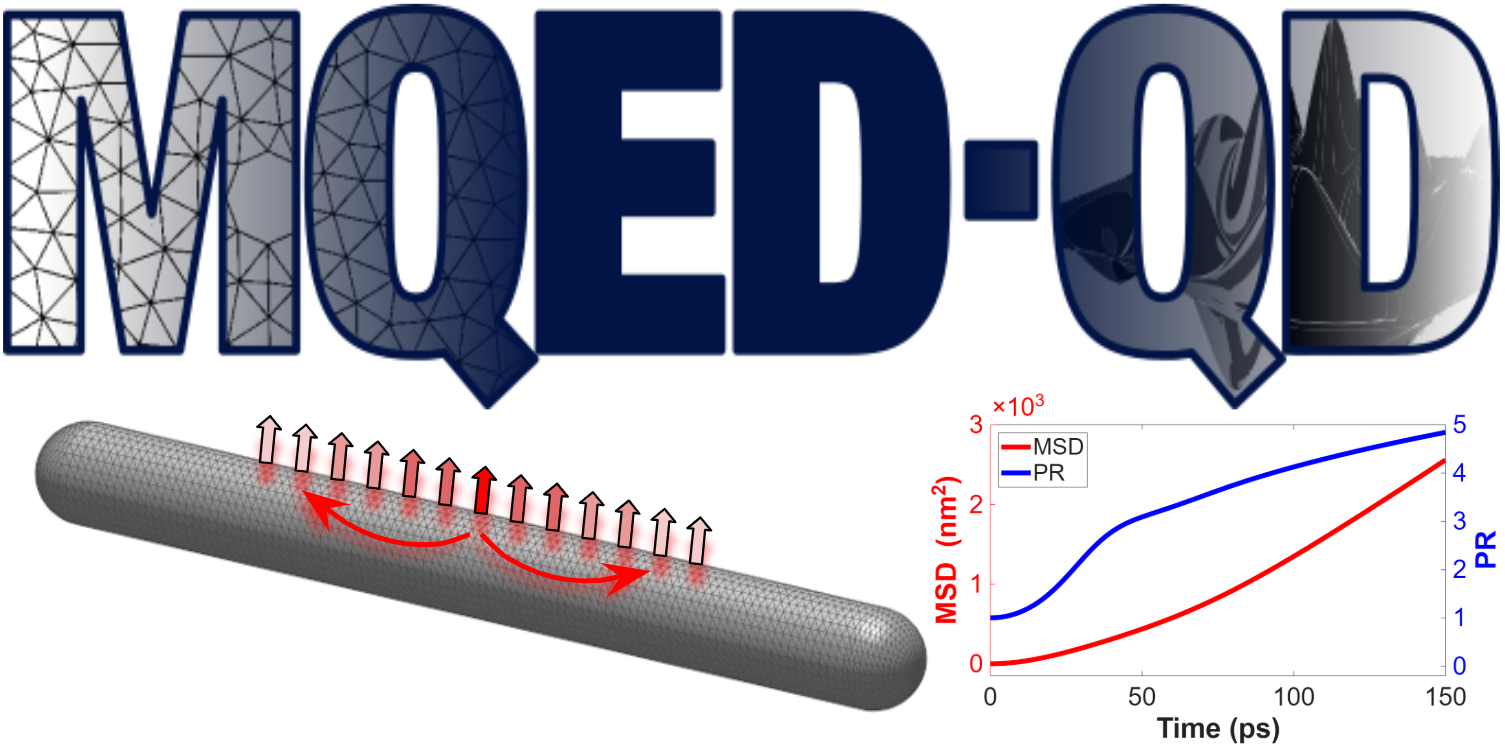}

\end{tocentry}

\begin{abstract}
Simulating the dynamics of molecular excitons in complex nanophotonic environments requires integrating rigorous electromagnetic simulations with accurate treatments of open quantum system dynamics. 
In this work, we develop \textsc{MQED-QD} (Macroscopic Quantum Electrodynamics for Quantum Dynamics), a robust computational package for simulating exciton dynamics in arbitrary dielectric and plasmonic environments. 
Based on the MQED framework, the package offers a unified workflow for constructing the dyadic Green's functions from classical electromagnetic solvers, parametrizing quantum master equations, and propagating the time evolution to determine the molecular subsystem's dynamical properties.
To demonstrate the package's capabilities, we simulate exciton transport within a one-dimensional molecular chain near a silver nanostructure, including benchmarking against planar surfaces and exploring the influence of silver nanorods.
Our results reveal that surface plasmon polaritons on nanorods dramatically enhance long-range dipole-dipole interactions, accelerating exciton delocalization and yielding higher participation ratios compared to planar geometries.
\textsc{MQED-QD} provides a powerful, open-source package that facilitates the rational design of nanoscale architectures by elucidating accurate molecular exciton dynamics in conjunction with nanophotonics and plasmonics.
\end{abstract}

\section{Introduction}

Exciton dynamics of molecular assemblies at the interface of plasmonic nanostructures has emerged as a central theme in organic optoelectronics, \cite{lee2024organic,che2024organic,martinez2021molecular,pei2020organic,zhang2019efficient,ostroverkhova2016organic,baldo1998highly}
light harvesting, \cite{singh2024degenerately,chen2022singlet,han2022integrating,yuan2022fluorescent,levi2015ultrafast,romero2014quantum,scholes2011lessons}
and quantum information science.\cite{scholes2025quantum,katsumi2025recent,harrington2022engineered,wehner2018quantum,heshami2016quantum,divincenzo1999quantum}
In such systems, the interaction between molecular excitation and the electromagnetic environment can profoundly modify energy transfer dynamics. For example, nearby metallic nanoparticles and thin dielectric films, resonance energy transfer (RET) can be enhanced or suppressed by orders of magnitude beyond the vacuum F\"orster limit \cite{forster1948zwischenmolekulare,bouchet2016long,hsu2017plasmon,dai2023phosphorescence} because the photonic density of states near a dielectric or metallic interface reshapes the dipole--dipole coupling strength that mediates exciton transport. \cite{dung2002intermolecular,wubs2004multiple,novotny2012principles,gonzaga-galeana_2017,cortes2020fundamental}
Furthermore, molecular polaritonics experiments have further shown that strong light--matter coupling in optical cavities creates hybrid exciton--photon states with dramatically altered transport properties.\cite{xiang2024molecular,xu_ultrafast_2023,garcia-vidal_manipulating_2021,herrera2020molecular,wang2020theory,hirai2020recent}
From theory and simulation perspectives, the key challenge is to treat the dynamics of a large ensemble of quantum emitters and their interactions with the dielectric environment on an equal footing.

The macroscopic quantum electrodynamics (MQED) theory, developed by Welsch and co-workers, offers an elegant framework by quantizing the electromagnetic field in the presence of arbitrary, dispersive, and absorbing media through the fluctuation--dissipation theorem.\cite{scheel2008macroscopic,buhmann2013dispersion}
In the MQED framework, all radiative inter-molecular coupling, dissipation rates, and energy level shifts can be expressed in terms of the classical dyadic Green's function,
which encodes the full influence of the dielectric environment. This feature has been applied to generalize F\"orster's $1/R^6$ energy transfer theory to include retardation, material dispersion, and geometric effects of the surrounding medium,\cite{dung2002intermolecular} and elucidate polariton-coupled electron transfer under the influence of dielectric environment. \cite{chuang2025quantum}
Despite much progress, these results almost always rely on high-symmetry geometries, in which the dyadic Green's function can be obtained analytically, such as Sommerfeld integrals for planar interfaces
or Mie theory for spheres. \cite{novotny2012principles,jia_calculation_2016,wu2018characteristic,lee2020controllable} 
For a nanostructure of arbitrary geometry or dispersion relation, one must resort to numerical methods such as the boundary element method (BEM) \cite{hohenester2012mnpbem}, the finite element method (FEM)\cite{Jin2015}
, and the finite-difference time-domain (FDTD) method.\cite{taflove_computational_2005}

Beyond the calculation of parameters related to electromagnetic fields, a second major challenge lies in integrating the MQED framework with the quantum dynamics of molecular emitters. The conventional approach utilizes semiclassical simulations that self-consistently combine Maxwell’s equations with Ehrenfest molecular dynamics\cite{ji_maxwelllink_2026,chen_ehrenfest+r_2019,li_necessary_2018} to capture rotation-vibration effects at complex plasmonic interfaces. \cite{sukharev_ultrafast_2014,sukharev_numerical_2014,sukharev_unveiling_2025} While recent quantum-corrected models have extended these classical simulations to incorporate nonlocal and tunneling effects,\cite{esteban2012bridging,savage2012revealing} the full MQED treatment of the field often leads to complex quantum dynamics problems characterized by long-range coupling, non-Markovian memory kernels, and generalized dissipation.\cite{chuang_2024} Capturing these dynamical effects accurately requires sophisticated methods, such as generalized quantum master equations or hierarchical equations of motion.\cite{campaioli2024quantum,dan2025simulating}Consequently, there remains an outstanding need for a platform that bridges numerically accurate Green's functions with efficient dynamical solvers for the open quantum system description of molecular excitons.

In this work, we developed \textsc{MQED-QD}, an open-source Python package to
(\emph{i}) extract the dyadic Green's function from classical electromagnetic solvers,
(\emph{ii}) construct the MQED Hamiltonian and the corresponding quantum master equation, and (\emph{iii}) propagate the quantum dynamics for coupled molecular emitters. 
The \textsc{MQED-QD} package allows for user-specified geometry and material parameters and facilitates the full pipeline for calculating the dyadic Green's functions by analytical solutions or numerical methods with a built-in calibration protocol.
Through interfacing with QuTiP,\cite{johansson2012qutip} the \textsc{MQED-QD} package simulates the dynamics of open quantum systems by integrating either the standard Lindblad master equation or a non-Hermitian Schr\"odinger equation.
Another key feature of the \textsc{MQED-QD} package is that the entire workflow is accessible through a set of command-line tools, lowering the barrier for non-specialist users.
For demonstration, we investigate exciton transport and delocalization properties of molecular aggregates in the presence of silver nanostructures of various geometries, including a planar surface and a nanorod.

The remainder of this paper is organized as follows. Section~\ref{sec:theoretical ramework} presents the MQED theoretical framework for coupled molecular excitons and dielectric environments.
Section~\ref{sec:Dyadic Green’s Function} describes the construction of the dyadic Green's function via analytical and numerical approaches.
Section~\ref{sec:Quantum Dynamics} provides an overview of the quantum dynamics propagation methods and the properties of exciton transport and delocalization. 
In Section~\ref{sec:Package Details}, we detail the package architecture and instructions for using the package.
In Section~\ref{sec:Results}, we demonstrate the robustness of the package through calibration benchmarks, exciton dynamics above a planar silver surface, and exciton transport near a silver nanorod.
We conclude in Section~\ref{sec:conclusion} with a summary and outlook.

\section{Theoretical Framework}\label{sec:theoretical ramework}

In this section, we provide an overview of the macroscopic quantum electrodynamics (MQED) and the quantum master equation approach for the dynamics of an ensemble of quantum emitters. For more detailed derivation and discussion, the interested reader is referred to Ref.~\citenum{hsu2025chemistry}.

Throughout the paper, we use the following notation. We use the bold face $\mathbf{r}$ for a vector in three-dimensional space and the double bar notation ($\overline{\overline{\mathbf{G}}}$) for a tensor. 
The inner product between two vectors ($\mathbf{a}$ and $\mathbf{b}$) is denoted by $\mathbf{a}\cdot\mathbf{b}$, and the dyadic product is denoted by $\mathbf{a}\mathbf{b}$ producing a tensor. 
The hat notation, such as $\hat{H}$, $\hat{\mathbf{E}}$, denotes a quantum operator. The subscript ($i,j$) denotes the Cartesian labels. The Greek letter subscript ($\alpha,\beta$) labels the emitters.

\subsection{The MQED Hamiltonian}

We consider an ensemble of $N_{\mathrm{M}}$ quantum subsystems coupled to electromagnetic fields in arbitrary dielectric environments. The total Hamiltonian takes the form $\hat{H}=\hat{H}_\mathrm{M}+\hat{H}_\mathrm{P}+\hat{V}_\mathrm{MP}$. 
We model each quantum subsystem by two quantum levels described by the ground state $|{\mathrm{g}}_{\alpha}\rangle$ and excited state $|{\mathrm{e}}_{\alpha}\rangle$ and write the matter Hamiltonian as 
\begin{align}
\hat{H}_\mathrm{M} = \sum_{\alpha=1}^{N_\mathrm{M}} \hbar \omega_\mathrm{M} \hat{\sigma}^{+}_\alpha \hat{\sigma}^{-}_\alpha
\label{Eq:H_M}
\end{align}
where $\hat{\sigma}^{+}_\alpha=|\mathrm{e}_{\alpha}\rangle\langle \mathrm{g}_{\alpha}|$ and $\hat{\sigma}^{-}_\alpha=|\mathrm{g}_{\alpha}\rangle\langle \mathrm{e}_{\alpha}|$ denote the creation and destruction operators for the $\alpha$-th subsystem.
Here we ignore the vibrational degrees of freedom in the quantum subsystem as we focus on the electronic transition at low temperature. 

In the framework of MQED, the Hamiltonian of quantized electromagnetic fields takes a similar form as the standard photon Hamiltonian
\begin{align}
\hat{H}_\mathrm{P} = \int d^{3}\mathbf{r}\int_{0}^{\infty}\, d\omega \, 
\hbar\omega \mathbf{\hat{f}^{\dagger}}(\mathbf{r},\omega) \mathbf{\hat{f}}(\mathbf{r},\omega)
\label{Eq:H_photon}
\end{align}
Here $\mathbf{\hat{f}^{\dagger}}(\mathbf{r},\omega)$ and $\mathbf{\hat{f}}(\mathbf{r},\omega)$ are the creation and annihilation operators of the quantized bosonic vector fields derived from the quantization of the hybrid medium excitations and photons,\cite{gruner_correlation_1995,dung_three-dimensional_1998} rather than photons in a vacuum. By design, the vector components of these quantized bosonic field operator satisfy the bosonic commutation relation, $[{\hat{f}_i}(\mathbf{r},\omega),{\hat{f}_j^{\dagger}}(\mathbf{r}',\omega')]=\delta_{ij}\delta(\mathbf{r}-\mathbf{r}')\delta(\omega-\omega')$ for $i,j\in\{x,y,z\}$. Namely, these vector field operators correspond to create and annihilate photons dressed by dielectric medium. 

In the multipolar gauge,\cite{feist_macroscopic_2021} the coupling between the quantum subsystems and the dressed photon field can be expressed in the form
\begin{align}
\hat{V}_\mathrm{MP} = 
-\sum_{\alpha=1}^{N_\mathrm{M}} ( \hat{\sigma}^{+}_\alpha+ \hat{\sigma}^{-}_\alpha)\boldsymbol{\mu}_{\alpha}\cdot\mathbf{\hat{E}}(\mathbf{r_{\alpha}}) ,
\label{Eq:V_MP}
\end{align}
The transition dipole moment of the $\alpha$-th subsystem is $\boldsymbol{\mu}_{\alpha}=\mu_\alpha\mathbf{n}_\alpha$ where $\mu_{\alpha}$ is its real-valued magnitude and $\mathbf{n}_{\alpha}$ is the unit vector of its direction. 
The electric field operator at position $\mathbf{r}_\alpha$ takes the form\cite{dung2002intermolecular}:
\begin{align}
\mathbf{\hat{E}}(\mathbf{r_{\alpha}}) = i\sqrt{\frac{\hbar}{\pi\epsilon_{0}}}\int d^{3}\mathbf{r_{\beta}}\int_{0}^{\infty}d\omega \frac{\omega^2}{c^2} 
\sqrt{\text{Im}[\epsilon_{\mathrm{r}}(\mathbf{r_{\beta}},\omega)]}\,\overline{\overline{\mathbf{G}}}(\mathbf{r_{\alpha}},\mathbf{r_{\beta}},\omega)\cdot\mathbf{\hat{f}} 
(\mathbf{r_{\beta}},\omega)+ \text{H.c}.
\label{Eq:field}
\end{align}
where $\epsilon_0$ is the vacuum permittivity, $\epsilon_\mathrm{r}(\mathbf{r_{\beta}},\omega)$ is the relative permittivity at position $\mathbf{r_{\beta}}$ and frequency $\omega$, and $c$ is the speed of light in vacuum. 
The tensor $\overline{\overline{\mathbf{G}}}(\mathbf{r_{\alpha}},\mathbf{r_{\beta}},\omega)$ denotes the dyadic Green's function satisfying the macroscopic Maxwell's equation in the frequency domain:
\begin{align}
\left( \frac{\omega^2}{c^2}\epsilon_\mathrm{r}(\mathbf{r}_\alpha,\omega) - \boldsymbol{\nabla} \times \boldsymbol{\nabla} \times \right)\overline{\overline{\mathbf{G}}}(\mathbf{r}_\alpha,\mathbf{r}_\beta,\omega) = 
- \delta(\mathbf{r}_\alpha-\mathbf{r}_\beta) \mathbf{\overline{\overline{I}}}_3,
\label{Eq:Green_Fun_full}
\end{align}
where $\mathbf{\overline{\overline{I}}}_3$ is the unit dyadic ($3\times 3$ identity tensor) and $\delta(\mathbf{r}_\alpha-\mathbf{r}_\beta)$ is a three dimensional delta function. 
We emphasize that the dyadic Green's function encodes the spatial propagation of the dressed photon field in arbitrary linear, inhomogeneous, dispersive and absorbing dielectric environments characterized by the relative permittivity $\epsilon_\mathrm{r}(\mathbf{r}_\alpha,\omega)$. Namely, the influence of dielectric environments on quantum subsystems can be quantitatively captured via the dyadic Green's function.

Since the macroscopic Maxwell's equation is linear, the dyadic Green's function can be decomposed into two contributions: 
\begin{align}
\overline{\overline{\mathbf{G}}}(\mathbf{r}_\alpha,\mathbf{r}_\beta,\omega) = \overline{\overline{\mathbf{G}}}_{0}(\mathbf{r}_\alpha,\mathbf{r}_\beta,\omega) + 
\overline{\overline{\mathbf{G}}}_{\mathrm{Sc}}(\mathbf{r}_\alpha,\mathbf{r}_\beta,\omega) ,
\label{Eq:Green_decomposed}
\end{align}
where $\overline{\overline{\mathbf{G}}}_0(\mathbf{r}_\alpha,\mathbf{r}_\beta,\omega)$ represent the vacuum dyadic Green's function and 
$\overline{\overline{\mathbf{G}}}_\mathrm{Sc}(\mathbf{r}_\alpha,\mathbf{r}_\beta,\omega)$ denotes the scattering Green's function which originates from the presence of dielectric environments. The vacuum dyadic Green's function can be expressed in a closed-form as:
\begin{align}
    \overline{\overline{\mathbf{G}}}_0(\mathbf{r}_\alpha,\mathbf{r}_\beta,\omega)=&
    \frac{e^{ik_0 R_{\alpha\beta}}}{4\pi R_{\alpha\beta}}
    \left\{\vphantom{\frac{e^R}{R}}
    \left(\overline{\overline{\mathbf{I}}}_3-{\mathbf{n}}_\mathrm{R}  {\mathbf{n}}_\mathrm{R} \right)\right. +\left.\left(3 {\mathbf{n}}_\mathrm{R}   {\mathbf{n}}_\mathrm{R}  
    -\overline{\overline{\mathbf{I}}}_3\right)\left[\frac{1}{(k_0 R_{\alpha\beta})^{2}}-\frac{i}{k_0 R_{\alpha\beta}}\right]
    \right\},
    \label{Eq:g0}
\end{align}
where $k_0 = \omega/c$ is the angular wavenumber in vacuum. The separation vector is defined by $\mathbf{r}_\alpha - \mathbf{r}_\beta =\mathbf{R}_{\alpha\beta} = R_{\alpha\beta} {\mathbf{n}}_\mathrm{R} $ where ${\mathbf{n}}_\mathrm{R}$ is the unit vector and  $R_{\alpha\beta}$ is the distance between the source and response points.
On the other hand, the scattering Green's function depends on the geometry of dielectric materials and their relative permittivity, which encodes the spatial and spectral properties of the dielectric environments. Thus, determining the scattering dyadic Green's function is almost always the key difficulty in the MQED framework.

\subsection{Quantum Dynamics of Molecular Exciton States }



To investigate the dynamics of the quantum subsystems interacting with the quantized electromagnetic fields, we then treat the hybrid system within the framework of open quantum systems. Starting from the Heisenberg equation of motion, we can partially trace out the continuum photonic degrees of freedom (photonic bath) and focus on the dynamics of reduced density matrix $\hat{\rho}_\mathrm{M}(t)=\mathrm{Tr}_\mathrm{P}[\hat{\rho}(t)]= \int d^3\mathbf{r}\int_0^\infty d\omega \hat{\rho}(\mathbf{r},\omega,t)$ where we assume the initial condition of the photonic bath is in the vacuum state. The Heisenberg equation can be reduced to a master equation with the generalized spectral density~\cite{sanchez2022few,chuang_2024,chuang_2024_2}:
\begin{align}
J_{\alpha\beta}(\omega)
= \frac{\omega^{2}}{\pi \hbar \epsilon_{0}c^2} 
\boldsymbol{\mu}_{\alpha}\cdot 
\mathrm{Im}\,\overline{\overline{\mathbf{G}}}
(\mathbf{r}_\alpha,\mathbf{r}_\beta,\omega)
\cdot 
\boldsymbol{\mu}_{\beta}.
\label{Eq:Gen_Spectral_Density}
\end{align}
This generalized spectral density characterizes the frequency-dependent coupling between two quantum subsystems ($\alpha$ and $\beta$) mediated by the electromagnetic environment, which is dictated by the dyadic Green's function. 

In this work, we focus on the weak light-matter coupling regime and assume that the generalized spectral density varies smoothly near the molecular transition frequency $\omega_\text{M}$.
Thus, we can evoke the Markov approximation and eliminate the spectral dependence of $J_{\alpha\beta}(\omega)$ by taking the two-point dyadic Green’s function as $\overline{\overline{\mathbf{G}}}(\mathbf{r}_\alpha,\mathbf{r}_\beta,\omega_{\mathrm{M}})$. Namely, the electromagnetic environment is considered a broadband Markovian reservoir with negligible memory effects.
Under this Markovian approximation, the resulting quantum master equation for the reduced density matrix takes the Lindbladian form\cite{chuang_2024_2,wang_robust_2025}
\begin{align}
\nonumber
\frac{\partial}{\partial t} \hat{\rho}_{\mathrm{M}}(t) = 
&-\frac{i}{\hbar} \left[ \hat{H}_{\mathrm{M}} + \hat{{H}}_\mathrm{CP} + \hat{{H}}_\mathrm{DDI}, \hat{\rho}_\mathrm{M}(t) \right] \\
&+ \sum_{\alpha,\beta}^{N_\mathrm{M}}  {\Gamma}_{\alpha\beta} \left( \hat{\sigma}_{\beta}^{-} \hat{\rho}_\mathrm{M}(t) \hat{\sigma}_{\alpha}^{+} - \frac{1}{2} \left\{ \hat{\sigma}_{\alpha}^{+} \hat{\sigma}_{\beta}^{-}, \hat{\rho}_\mathrm{M}(t) \right\} \right),
\label{Eq:QME}
\end{align}
where $[\hat{O}_1, \hat{O}_2]= \hat{O}_1  \hat{O}_2 - \hat{O}_2\hat{O}_1 $ and $\{ \hat{O}_1, \hat{O}_2 \} = \hat{O}_1  \hat{O}_2 + \hat{O}_2\hat{O}_1 $ are the commutator and the anti-commutator of two operators, respectively.
Here, $\hat{H}_\mathrm{CP}$ and $\hat{H}_\mathrm{DDI}$ represent
the Casimir-Polder (CP) potential and the dipole-dipole interactions (DDI), respectively\cite{chuang_2024,chuang_2024_2, wang_robust_2025}.
\begin{align}
\label{Eq:DDI_Ham} & \hat{{H}}_\mathrm{DDI} =  \sum_{\alpha,\beta (\alpha\neq \beta) }^{N_\mathrm{M}} V_{\alpha\beta} \, \hat{\sigma}^{+}_\alpha \hat{\sigma}^{-}_\beta , \\
&\hat{{H}}_\mathrm{CP} = \sum_{\alpha=1}^{N_\mathrm{M}} \Lambda_{\alpha}^{\text{Sc}} \, \hat{\sigma}^{+}_\alpha \hat{\sigma}^{-}_\alpha .
\label{Eq:H_LS}
\end{align}
The DDI strength $V_{\alpha\beta}$ in eq~\ref{Eq:DDI_Ham} can be evaluated using the real part of the two-point dyadic Green's function:
\begin{equation}
\label{Eq:DDI}
V_{\alpha\beta} = \frac{- \omega_\mathrm{M}^2}{ \epsilon_0 c^2}  \boldsymbol{\mu}_{\alpha} \cdot \mathrm{Re} \overline{\overline{\mathbf{G}}}(\mathbf{r}_\alpha,\mathbf{r}_\beta,\omega_\mathrm{M}) \cdot \boldsymbol{\mu}_{\beta},
\end{equation}
and the CP potential $\Lambda_{\alpha}^{\mathrm{Sc}}$ describes the transition energy shift induced by the dielectric environment, which can be evaluated by:\cite{chuang_2024,wang_robust_2025}  
\begin{align}
\Lambda_{\alpha}^\mathrm{Sc} =& \mathcal{P} \int_{0}^{\infty} d\omega \frac{\omega^2}{\pi \varepsilon_0 c^2}   \left( \frac{1}{ \omega + \omega_\mathrm{M} } -\frac{1}{\omega - \omega_\mathrm{M} } \right) {\boldsymbol{\mu}_{\alpha} \cdot \mathrm{Im}\overline{\overline{\mathbf{G}}}_{\text{Sc}}(\mathbf{r}_\alpha,\mathbf{r}_\alpha,\omega) \cdot \boldsymbol{\mu}_{\alpha}} , \label{Eq:Interaction_Definition}
\end{align}
where $\mathcal{P}$ denotes the Cauchy principal value. 
The generalized dissipation rate in eq~\ref{Eq:QME} is given by the imaginary part of the two-point dyadic Green's function:
\begin{equation}
\label{Eq:Gamma}
\Gamma_{\alpha\beta} = \frac{2\omega_{\mathrm{M}}^2}{\hbar\varepsilon_0 c^2} \boldsymbol{\mu}_{\alpha} \cdot \mathrm{Im}\overline{\overline{\mathbf{G}}}(\mathbf{r}_{\alpha},\mathbf{r}_{\beta},\omega_{\mathrm{M}}) \cdot \boldsymbol{\mu}_{\beta}.    
\end{equation}
Note that, in the weak coupling limit, the diagonal term of the dissipation matrix $\Gamma_{\alpha\alpha}$ describes the population decay of the electronic excited state of a single molecule in a complex dielectric environment.\cite{dung_spontaneous_2000}. The ratio between the dissipation rate ($\Gamma_{\alpha\alpha}$) and the spontaneous emission rate ($\Gamma_0$) corresponds to the Purcell enhancement factor $\Gamma_{\alpha\alpha}/\Gamma_0=\frac{6\pi c}{\omega_\mathrm{M}}\mathbf{n}_{\alpha} \cdot \mathrm{Im}\overline{\overline{\mathbf{G}}}(\mathbf{r}_{\alpha},\mathbf{r}_{\alpha},\omega_{\mathrm{M}}) \cdot \mathbf{n}_{\alpha}$.\cite{purcell_resonance_1946} Thus, the single-point dyadic Green’s function can be evaluated in terms of the Purcell factor, which is available in classical electrodynamics simulations.


\subsection{Exciton Transport and Delocalization}
Within this MQED framework, we characterize exciton transport and delocalization using two key observables: the mean-square displacement (MSD) and the participation ratio (PR).\cite{murphy2011generalized,sokolovskii2025strong} 
Given a single excitation initially localized at site $\alpha_0$, the MSD quantifies the spatial spread of the exciton from the initial site, while the PR measures the effective number of sites participating in the quantum state. The formal definitions of MSD and PR, including the renormalization procedure required for dissipative dynamics, are presented in Section~\ref{sec:Quantum Dynamics}. 
In the following sections, we demonstrate the robustness of the \textsc{MQED-QD} package by benchmarking the MSD and PR in complex dielectric environments. 
Note that, since we integrate the quantum master equation for the time evolution of the reduced density matrix, the current framework is not limited to these observables.

\section{Construction of the Dyadic Green’s Function}\label{sec:Dyadic Green’s Function}

The key challenge in the MQED framework is to construct the two-point dyadic Green's function, especially for a dielectric environment with arbitrary geometric structures.
In this section, we present a practical approach to constructing the dyadic Green's function using existing computational electrodynamic tools.

\subsection{Analytical Solutions}
For specific geometric structures, such as an isotropic sphere and an infinite planar surface, we can employ the analytical solutions to Maxwell's equations in terms of spectral expansion. The scattering of electromagnetic waves by an isotropic, spherical particle can be evaluated as a sum of spherical harmonic functions using Mie theory,\cite{jia_calculation_2016} which can be used for calculating the dyadic Green's functions.\cite{lee2020controllable}
For a single point source over a planar layered dielectrics, one can expand the spherical wave in terms of sum of plane waves and evanescent waves using Fourier transformation in space, known as the Sommerfeld identity.\cite{jackson_classical_2009} The reflection component of the dyadic Green's function can be directly evaluated using the Fresnel reflection coefficients.\cite{wu2018characteristic}

\subsection{Numerical Methods}
Despite these existing analytical solutions, for more complex material geometries, one must employ numerical solvers for macroscopic Maxwell's equations. Many computational electrodynamics simulation schemes have been developed, including the finite-difference time-domain (FDTD) approach, finite-element method (FEM), and boundary-element method (BEM). These numerical solvers yield the classical electric field vector, rather than the tensor element of the dyadic Green's function required in eq~\ref{Eq:Green_Fun_full}. 

To construct the dyadic Green's function, we utilize the fundamental relationship between an oscillating source point dipole and the electric fields it generates\cite{somayaji2025remarkable,hsu2025chemistry,ding2017plasmon}.
\begin{align}
{E}_{i}(\mathbf{r}_\alpha,\omega_\mathrm{M}) = \frac{\omega^2}{\epsilon_0 c^2} \sum_{j=1}^3 {G}_{ij}(\mathbf{r}_\alpha,\mathbf{r}_\beta,\omega_\mathrm{M}) {p}_{j}(\omega_\mathrm{M}) .
\label{Eq:GF_field}
\end{align}
Here, $E_i(\mathbf{r}_\alpha,\omega_\mathrm{M})$ and $p_j(\omega_\mathrm{M})$ denote the components of and the electric field vector $\mathbf{E}(\mathbf{r}_\alpha,\omega_\mathrm{M})$ calcuated at position $\mathbf{r}_\alpha$ and the source dipole moment $\mathbf{p}(\omega_\mathrm{M})$ oscillating at frequency $\omega_\mathrm{M}$ located at position $\mathbf{r}_\beta$, respectively.
To evaluate the specific matrix elements $G_{ij}(\mathbf{r}_\alpha, \mathbf{r}_\beta, \omega_\mathrm{M})$ of the dyadic Green's function, we perform a set of calculations for the source dipole moment chosen to be in the Cartesian axis $j \in \{x, y, z\}$ and calculate the $i$-component of the electric field $E_i(\mathbf{r}_\alpha,\omega_\mathrm{M})$ by solving the frequency-domain Maxwell's equations. For each calculation, we can determine the individual tensor elements by
\begin{equation}
    G_{ij}(\mathbf{r}_\alpha, \mathbf{r}_\beta, \omega_\mathrm{M}) = \frac{\epsilon_0 c^2}{\omega^2} \frac{E_i(\mathbf{r}_\alpha, \omega_\mathrm{M})}{p_j(\omega_\mathrm{M})}.
    \label{Eq:GF_reconstruction_general}
\end{equation}
For any two points in space, this procedure requires three independent simulations to determine all components of the dyadic Green's function.

\subsubsection{Boundary Element Method (BEM)}
We employ the boundary element method (BEM)\cite{garcia_de_abajo_retarded_2002}, specifically the MNPBEM toolbox designed for the simulation of metallic nanoparticles (MNP).\cite{hohenester2012mnpbem}
In contrast to general volume-based simulation tools, such as FDTD, the BEM approach discretizes material interfaces and reformulates Maxwell’s equations into surface integral equations. The dielectric environment is then described through local and isotropic dielectric functions $\epsilon_\lambda(\omega)$ within each domain (labeled by $\lambda$) separated by well-defined boundaries.
Consequently, by restricting discretization to the interfaces rather than the entire volume, the BEM significantly reduces the degrees of freedom and the computational cost compared to volume-based methods.

\subsubsection{BEM Calibration Protocol}
\label{sec:calibration_dipole}
To incorporate the MNPBEM toolbox into the calculation of the two-point dyadic Green's functions for the MQED framework, we design a calibration protocol to ensure the unit is consistent with SI units. 
Because the electric fields and dipole moments in the MNPBEM toolbox are expressed in solver-specific internal units rather than SI units, we determine the effective dipole normalization factor from vacuum calculations and subsequently apply this factor to reconstruct the dyadic Green’s function using eq~\ref{Eq:GF_reconstruction_general}.

The detailed calibration procedures of the dipole intensity are as follows:
\begin{enumerate}
    \item \textbf{Calculating analytical reference field in vacuum.}
    We first place an oscillating point dipole of known SI magnitude $p_j(\omega_{\mathrm M}) = 1~\mathrm{C\cdot m}$ at position $\mathbf r_\beta$ in vacuum and compute the corresponding electric field component $E_{0,z}(\mathbf r_\alpha,\omega_{\mathrm M})$ at a set of observation points $\{\mathbf r_\alpha\}$ using the
    analytical vacuum Green's tensor $G_{0,zj}$ in eq~\ref{Eq:g0}]:
    \begin{equation}
        E_{0,z}(\mathbf r_\alpha,\omega_{\mathrm M}) = \frac{\omega_{\mathrm M}^2}{\epsilon_0 c^2}\,
        G_{0,zj}(\mathbf r_\alpha,\mathbf r_\beta,\omega_{\mathrm M}) p_j(\omega_{\mathrm M}).
        \label{Eq:E0_analytic}
    \end{equation}

    \item \textbf{Calculating vacuum BEM field for a unit source.}
    Next, we perform an equivalent vacuum simulation using the MNPBEM toolbox. Using the same observation points $\{\mathbf r_\alpha\}$ and frequency $\omega_{\mathrm M}$, we initialize the solver with a unit BEM dipole $p_j^{\mathrm{BEM}}(\omega_{\mathrm M})=1$ BEM unit oriented along direction $i$ at $\mathbf r_\beta$.  The solver returns the corresponding field $E_{0,z}^{\mathrm{BEM}}(\mathbf r_\alpha,\omega_{\mathrm M})$, which may be written as
    \begin{equation}
        E_{0,z}^{\mathrm{BEM}}(\mathbf r_\alpha,\omega_{\mathrm M})
        = \frac{\omega_{\mathrm M}^2}{\epsilon_0 c^2} G_{0,zj}^{\mathrm{BEM}}(\mathbf r_\alpha,\mathbf r_\beta,\omega_{\mathrm M})p_j^{\mathrm{BEM}}(\omega_{\mathrm M}).
        \label{Eq:E0_BEM}
    \end{equation}

    \item \textbf{Evaluating the calibration coefficient using least-squares fitting.}
    We then determine a complex scalar $s(\omega_{\mathrm M})$ that minimizes the squared norm difference between the analytical field and the BEM field, i.e.
    \begin{equation}
        s(\omega_{\mathrm M})
        =
        \arg\min_{s}\left[
        \sum_{\alpha}
        \left|
        E_{0,z}^{\mathrm{BEM}}(\mathbf r_\alpha,\omega_{\mathrm M})
        -
        s\,E_{0,z}(\mathbf r_\alpha,\omega_{\mathrm M})
        \right|^2\right]
        \label{Eq:LS_calibration}
    \end{equation}
    In principle, the BEM field in vacuum should differ from the analytical reference only by a global complex scaling factor. In practice, we exclude observation points closer than a minimum distance $d_\mathrm{min}$ from the dipole to avoid near-source discretization artifacts
    (see Figure~\ref{Fig:calibration}). 
    We define the exclusive summation as $\sum^\prime_\alpha=\sum_{|\mathbf{r}_\alpha-\mathbf{r}_\beta|>d_\mathrm{min}}$, and the complex scaling factor has the closed-form expression
    \begin{equation}
        s(\omega_{\mathrm M})=
        \frac{\sum_\alpha^{\prime}
              E_{0,z}^*(\mathbf r_\alpha,\omega_{\mathrm M})
              E_{0,z}^{\mathrm{BEM}}(\mathbf r_\alpha,\omega_{\mathrm M})}
             {\sum_\alpha^{\prime}
              \left|E_{0,z}(\mathbf r_\alpha,\omega_{\mathrm M})\right|^2},
              \label{eq:s_closedform}
    \end{equation}
    which corresponds to the complex least-squares fitting implemented in the \textsc{MQED-QD} package.
    Note that, if we divided eq~\ref{Eq:E0_BEM} by $s(\omega_{\mathrm{M}})$, we recover $E_{0.z}(\mathbf r_\alpha,\omega_{\mathrm M})$ by eq~\ref{Eq:E0_analytic} in SI unit 
    \begin{equation}
        \frac{E_{0,z}^{\mathrm{BEM}}(\mathbf r_\alpha,\omega_{\mathrm M})}{s(\omega_{\mathrm{M}})}
        =
        \frac{\omega_{\mathrm M}^2}{\epsilon_0 c^2}\,
        G_{0,zj}^{\mathrm{BEM}}(\mathbf r_\alpha,\mathbf r_\beta,\omega_{\mathrm M})\,
        \frac{p_j^{\mathrm{BEM}}(\omega_{\mathrm M})}{s(\omega_{\mathrm{M}})}\approx E_{0.z}(\mathbf r_\alpha,\omega_{\mathrm M}),
        \label{Eq:E0_divide_s}
    \end{equation}
    suggesting that $\frac{E_{0,z}^{\mathrm{BEM}}(\mathbf r_\alpha,\omega_{\mathrm M})}{s(\omega)}$ gives the vacuum field in SI unit.
    Comparing eq~\ref{Eq:E0_analytic} and eq~\ref{Eq:E0_divide_s}, we can find the following mapping $G_{0,zi} = G_{0,zi}^\mathrm{BEM}$, $p_i(\omega_\mathrm{M}) = \frac{p_i^\mathrm{BEM}}{s}$, and $ E_{0,z}=\frac{E_{0,z}^{\mathrm{BEM}}}{s}$. In other words, setting a source dipole of magnitude $\tilde{p}_i(\omega_\mathrm{M})=\frac{p_i^{\mathrm{BEM}}(\omega_{\mathrm M})}{s(\omega_{\mathrm{M}})}$ produces the same electric field as a dipole of magnitude $p_i(\omega_{\mathrm{M}}) = 1~\mathrm{C \cdot m}$ in SI unit.
\end{enumerate}

After this calibration process for points of interest for our simulation, the electric field generated by the BEM calculation for an arbitrary dielectric environment follows the relation
\begin{equation}
        \frac{E_{i}^{\mathrm{BEM}}(\mathbf r_\alpha,\omega_{\mathrm M})}{s(\omega_{\mathrm{M}})}
        =
        \frac{\omega_{\mathrm M}^2}{\epsilon_0 c^2}\,
        G_{ij}^{\mathrm{BEM}}(\mathbf r_\alpha,\mathbf r_\beta,\omega_{\mathrm M})\,
        \tilde{p}_j(\omega_\mathrm{M}),
        \label{Eq:E_over_s}
\end{equation}
where $\tilde{p}_j(\omega_\mathrm{M})$ is a unit dipole moment given in SI unit. Therefore, eq.~\ref{Eq:E_over_s} is our working equation to construct the tensor element of the dyadic Green's function. For any two points in space, we need to perform three BEM simulations with a unit-amplitude point dipole oriented along $j\in\{x,y,z\}$ and record the induced electric-field components $E_i^{\mathrm{BEM}}(\mathbf r_\alpha,\omega_\mathrm{M})$ for $i\in\{x,y,z\}$. 

Note that this calibration coefficient depends on the oscillating frequency of the source dipole $\omega_\mathrm{M}$, but not on the surrounding dielectric environment. Thus, $s(\omega_{\mathrm{M}})$ as obtained in vacuum can be used to convert the BEM fields for constructing the dyadic Green's functions for an arbitrary dielectric environment.

\section{Propagation of Quantum Dynamics}\label{sec:Quantum Dynamics}
To simulate the dynamics of the quantum subsystem, we focus on the single-excitation manifold. In the the excitation site basis, we define the overall ground state as $|\mathrm{G}\rangle=\prod_{\alpha=1}^{N_\mathrm{M}}|\mathrm{g}_\alpha\rangle$ and the single excitation state as $|\mathrm{X}_{\alpha}\rangle=|\mathrm{e}_\alpha\rangle\prod_{\alpha'\neq\alpha}^{N_\mathrm{M}}|\mathrm{g}_{\alpha'}\rangle$ (where the $\alpha$-th molecule is excited and all others are in the ground state).
Within this basis, we can express $\hat{\sigma}_{\alpha}^+=|\mathrm{X}_{\alpha}\rangle\langle\mathrm{G}|$ and $\hat{\sigma}_{\beta}^-=|\mathrm{G}\rangle\langle\mathrm{X}_{\beta}|$.
The matrix form of the matter Hamiltonian and the DDI coupling Hamiltonian can be expressed as   $\hat{H}_{\mathrm{M}}=\sum_{\alpha}\hbar\omega_\mathrm{M}|\mathrm{X}_{\alpha}\rangle\langle\mathrm{X}_{\alpha}|$ and 
$\hat{H}_{\mathrm{DDI}}=\sum_{\alpha\neq\beta}V_{\alpha\beta}|\mathrm{X}_{\alpha}\rangle\langle\mathrm{X}_{\beta}|$ respectively.


\subsection{Integrating Lindblad Master Equation}

We can reduce the generalized Lindblad master equation (eq~\ref{Eq:QME}) to the standard Lindblad form for the density matrix propagation. First, to avoid numerical issues, we enforce the dissipation matrix to be positive semidefinite (i.e. ensuring all eigenvalues are non-negative and real values) by symmetrization:\cite{wang_theory_2024}
$\bar{\Gamma}_{\alpha\beta}=\frac{1}{2}({\Gamma}_{\alpha\beta}+{\Gamma}_{\beta\alpha}^*)$.
Next, we diagonalize the symmetrized matrix $[\bar{\Gamma}_{\alpha\beta}]$ to obtain eigenvalues $\gamma_\alpha$ and the unitary transformation $U_{\alpha\beta}$ 
(i.e. $\bar{\Gamma}_{\alpha\beta}=\sum_{\alpha'}U_{\alpha\alpha'}\gamma_{\alpha'}U_{\beta\alpha'}^*$), so that we can define the jump operators by
$\hat{L}_{\alpha}=\sum_{\alpha'} U_{\alpha'\alpha}^*\hat{\sigma}_{\alpha'}=\sum_{\alpha'} U_{\alpha'\alpha}^*|\mathrm{G}\rangle\langle\mathrm{X}_{\alpha'}|$.
In the end, the standard form of the Lindblad master equation reads
\begin{equation}
\frac{\partial}{\partial t}\hat{\rho}_{\mathrm{M}}
= -\frac{i}{\hbar}\bigl[\hat{H}_{\mathrm{M}} + \hat{{H}}_\mathrm{DDI}, \hat{\rho}_{\mathrm{M}}\bigr]
+ \sum_{\alpha=1}^{N_{\mathrm{M}}} \gamma_{\alpha} \left(
\hat{L}_{\alpha} \hat{\rho}_{\mathrm{M}} \hat{L}_{\alpha}^\dagger
- \frac{1}{2}\left\{ \hat{L}_{\alpha}^\dagger \hat{L}_{\alpha}, \hat{\rho}_{\mathrm{M}} \right\}
\right),
\label{Eq:Lindblad_standard}
\end{equation}
With the standard Lindblad form, we can propagate the density matrix using the QuTiP toolbox \texttt{mesolver}.

\subsection{Non-Hermitian Hamiltonian Approach}
Since we focus on the dynamics of excitation states, we can employ an effective non-Hermitian Hamiltonian approach to further simplify the calculation.\cite {celardo2012superradiance,giusteri_non-hermitian_2015,chavez_disorder-enhanced_2021}
Instead of propagating the reduced density matrix in eq~\ref{Eq:QME}, we consider the state vector of the quantum subsystems restricted in the single-excitation manifold, $|\Psi(t)\rangle =  \sum_{\alpha=1}^{N_\mathrm{M}} c_\alpha(t)|\mathrm{X}_{\alpha}\rangle$, following the Schr\"odinger equation $i\hbar\frac{\partial}{\partial t}|\Psi(t)\rangle = \hat{H}_\mathrm{eff}|\Psi(t)\rangle$ with the effective non-Hermitian Hamiltonian defined as
\begin{align}
\hat{H}_{\mathrm{eff}}
&=
\hat{H}_\mathrm{M}+\hat{H}_\mathrm{DDI}- \dfrac{i}{2}\sum_{\alpha,\beta}\bar{\Gamma}_{\alpha\beta}|\mathrm{X}_{\alpha}\rangle\langle\mathrm{X}_{\beta}|\\
&=
\hat{H}_\mathrm{M}+\hat{H}_\mathrm{DDI}- \dfrac{i}{2}\sum_{\alpha}{\gamma}_{\alpha}\hat{L}_\alpha^\dagger\hat{L}_\alpha
\label{Eq:H_eff}
\end{align}
Here, the unitary transformation for the Lindblad operator is applicable to diagonalize the non-Hermitian part. Note that the non-Hermitian Hamiltonian drops the quantum jump term (i.e., the excitation does not jump back to the ground state). Consequently, the time evolution of the quantum state does not conserve the population and cannot capture an observable involving the ground state.
In other words, the population loss of $|\Psi(t)\rangle$ reflects irreversible population decay out of the single-excitation
manifold. 

\subsection{Observables}

To evaluate the MSD and PR as functions of time, we first compute the site populations $P_\alpha(t)=\langle\mathrm{X}_\alpha|\hat{\rho}_{\mathrm{M}}(t)|\mathrm{X}_\alpha\rangle=|c_\alpha(t)|^2$. 
Because dissipation induced by the dielectric environment causes the total population of the reduced density matrix to decay, we define the renormalized populations 
\begin{equation}
\tilde{P}_\alpha(t)=\frac{P_\alpha(t)}{\sum_{\beta=1}^{N_\mathrm{M}}P_\beta(t)} \, .
\label{Eq:Pop_renorm}
\end{equation}
The MSD relative to the initially excited site $\alpha_0$ is then 
\begin{equation}
\mathrm{MSD}(t)=\sum_{\alpha=1}^{N_{\mathrm{M}}} \tilde P_\alpha(t)\,
\bigl|\mathbf r_\alpha-\mathbf r_{\alpha_0}\bigr|^2 ,
\label{Eq:MSD}
\end{equation}
and the participation ratio (PR) is defined as
\begin{equation}
\mathrm{PR}(t)=\frac{1}{\sum_{\alpha=1}^{N_{\mathrm{M}}} [\tilde{P}_\alpha(t)]^2} \, .
\label{Eq:PR}
\end{equation}
As a final note, our implementation of the MQED-QD package allows for the evaluation of any operator in the single-excitation manifold of the general form $\hat{O}=\sum_{\alpha\beta} O_{\alpha\beta} |\mathrm{X}_\alpha\rangle\langle\mathrm{X}_\beta|$.

\section{MQED-QD Package Details}\label{sec:Package Details}

\begin{table}
  \caption{Example commands to use the MQED-QD package}
  \label{tbl:pkg_example}
  \begin{tblr}{ll}
    \hline
    command & description  \\
    \hline
    \hline
    \texttt{mqed\_GF\_Sommerfeld} & Calculate the dyadic Green's function using \\ & Sommerfeld integrals for planar surface \\
    \texttt{mqed\_BEM\_reconstruct\_GF} & Reconstruct dyadic Green's function through BEM results\\
    \hline[dashed]
    \texttt{mqed\_FE} & Calculate DDI ratio by electric field enhancement \\
    \texttt{mqed\_lindblad} & Simulate Lindblad master equation (eq~\ref{Eq:QME})  \\
    \texttt{mqed\_nhse}  & Simulate effective non-Hermitian Schrödinger equation (eq~\ref{Eq:H_eff}) \\
    \hline[dashed]
    \texttt{mqed\_plot\_msd} & Plot mean squared displacement (MSD)  after \\ & collecting results.\\
    \texttt{mqed\_plot\_PR} & Plot participation ratio (PR) after collecting results.\\
    \hline
  \end{tblr}
\end{table}

\begin{figure}[h!]
    \centering
    \includegraphics[width=0.95\linewidth]{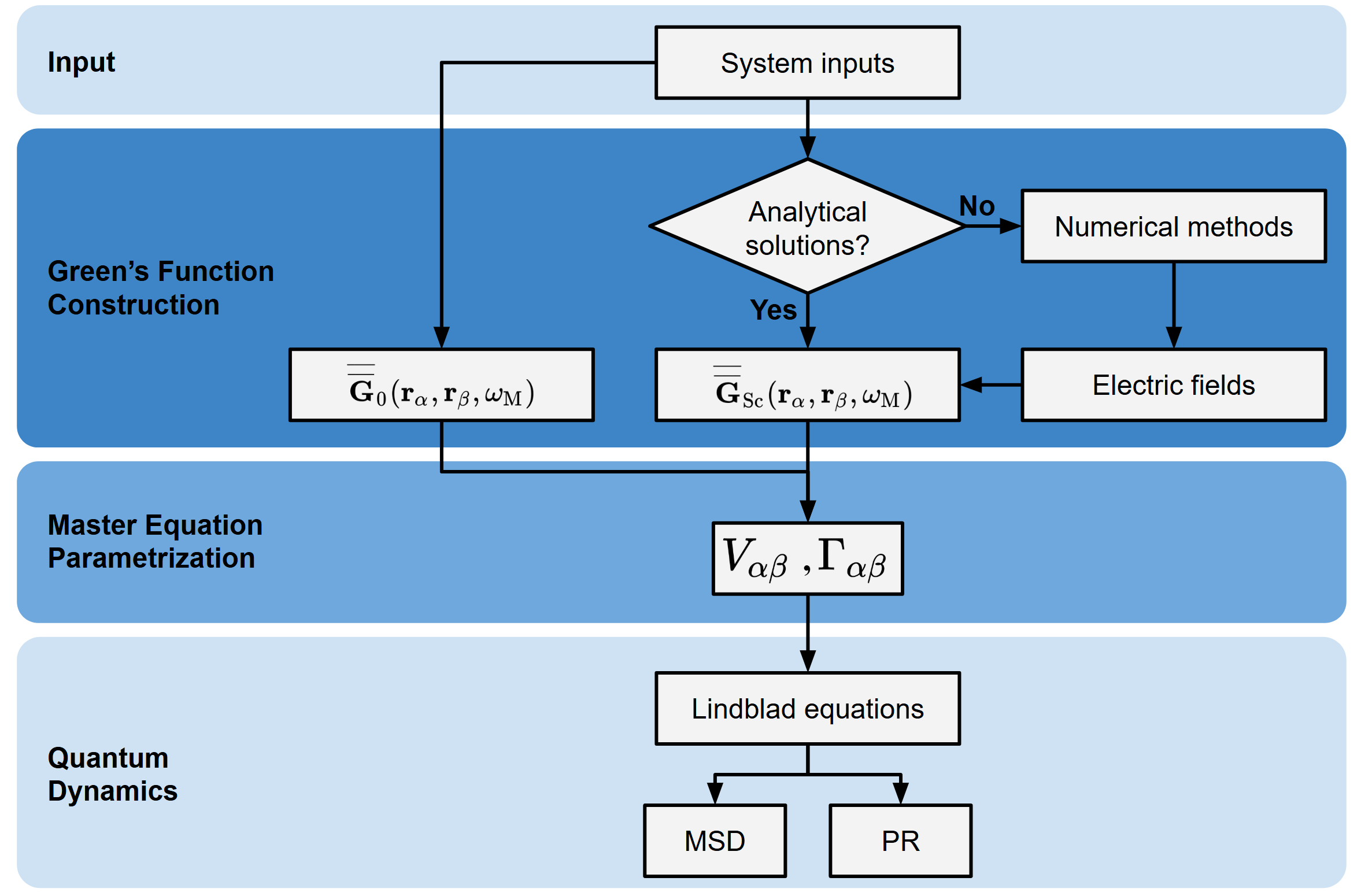}
    \caption{The workflow of the \textsc{MQED-QD} package to investigate exciton transport in the presence of dielectric environments. }
    \label{Fig:workflow}
\end{figure}

The primary outcome of this work is the open-source Python package (\textsc{MQED-QD}), providing an end-to-end workflow for quantum dynamics in dielectric environments. 
The package is designed to be user-friendly and to extend its functionalities systematically. The overall workflow is summarized in Figure~\ref{Fig:workflow} and comprises the following parts.
\begin{itemize}
    \item \textbf{System Input:} The simulation parameters are specified in a YAML configuration file and parsed through the Hydra framework, which also stores the full input configuration and outputs in a time-stamped directory to facilitate reproducibility. 
    \item \textbf{Green's Function Construction:} The \textsc{MQED-QD} package computes the dyadic Green’s functions as a tensor for selected positions through either (\emph{i}) the analytical solution, such as the Sommerfeld integral formulations for planar interfaces, or (\emph{ii}) the BEM fields for an arbitrary dielectric environment. The calibration protocol is implemented for a given frequency $\omega_\mathrm{M}$ and selected positions $\{\mathbf{r}_\alpha\}$. 
    \item \textbf{Master Equation Parametrization:} The DDI strength $V_{\alpha\beta}$ and the dissipation rate matrix $\Gamma_{\alpha\beta}$ are computed using eq~\ref{Eq:DDI} and eq~\ref{Eq:Gamma}, repsctively. Here the Hamiltonian of eq \ref{Eq:QME} are constructed in atomic unit (a.u.), and all the unit conversion are handled internally in the package. 
    \item \textbf{Quantum Dynamics Simulation:} Using these electromagnetic inputs, \textsc{MQED-QD} supports quantum-dynamics simulations based on the Lindblad master equation (eq~\ref{Eq:Lindblad_standard}) or, when appropriate, the Schr\"odinger equation with the effective non-Hermitian Hamiltonian (eq~\ref{Eq:H_eff}). Time evolution of the open quantum system dynamics is propagated using QuTiP. Transport and delocalization observables, including the mean-square displacement (eq~\ref{Eq:MSD}) and participation ratio (eq~\ref{Eq:PR}), are computed for output. 
\end{itemize}

The representative command-line entry points are listed in Table~\ref{tbl:pkg_example}. 
The MQED-QD package is openly accessible on our Github repository\cite{GitHub-repo} and supports the standard Python package installer (PIP).

\section{Results and Discussion}\label{sec:Results}

\subsection{Simulation Details}
To demonstrate the robustness of the workflow, we consider a chain of molecular emitters in close proximity to a silver nanostructure.
To be consistent with the experimental reports and common practice in the MQED literature, we evaluate the dyadic Green’s functions in SI units using either the analytical solution or the calibrated BEM calculations. The dielectric function of silver nanostructure is given by the MNPBEM toolbox.\cite{hohenester2012mnpbem} 
The \textsc{MQED-QD} package allows users to define the dielectric functions as an input file \texttt{dielectric.xlxs}. 

The molecular transition energy is chosen to be $\omega_\mathrm{M}= 1.864~\text{eV}$ (corresponding to the wavelength $\lambda_\mathrm{M}=665~\text{nm}$). Each molecule has a transition dipole moment of $\mu_\alpha = 3.8~\text{Debye}$, oriented along the $z$ direction (normal to the surface). Time evolution of the reduced density matrix $\hat{\rho}_\mathrm{M}(t)$ or the quantum state vector $|\Psi(t)\rangle$ is performed using the QuTiP tool with adaptive time step for convergence. The observable operators are evaluated every $5~\text{fs}$.

\begin{figure}[htbp]
    \centering
    \includegraphics[width=0.5\linewidth]{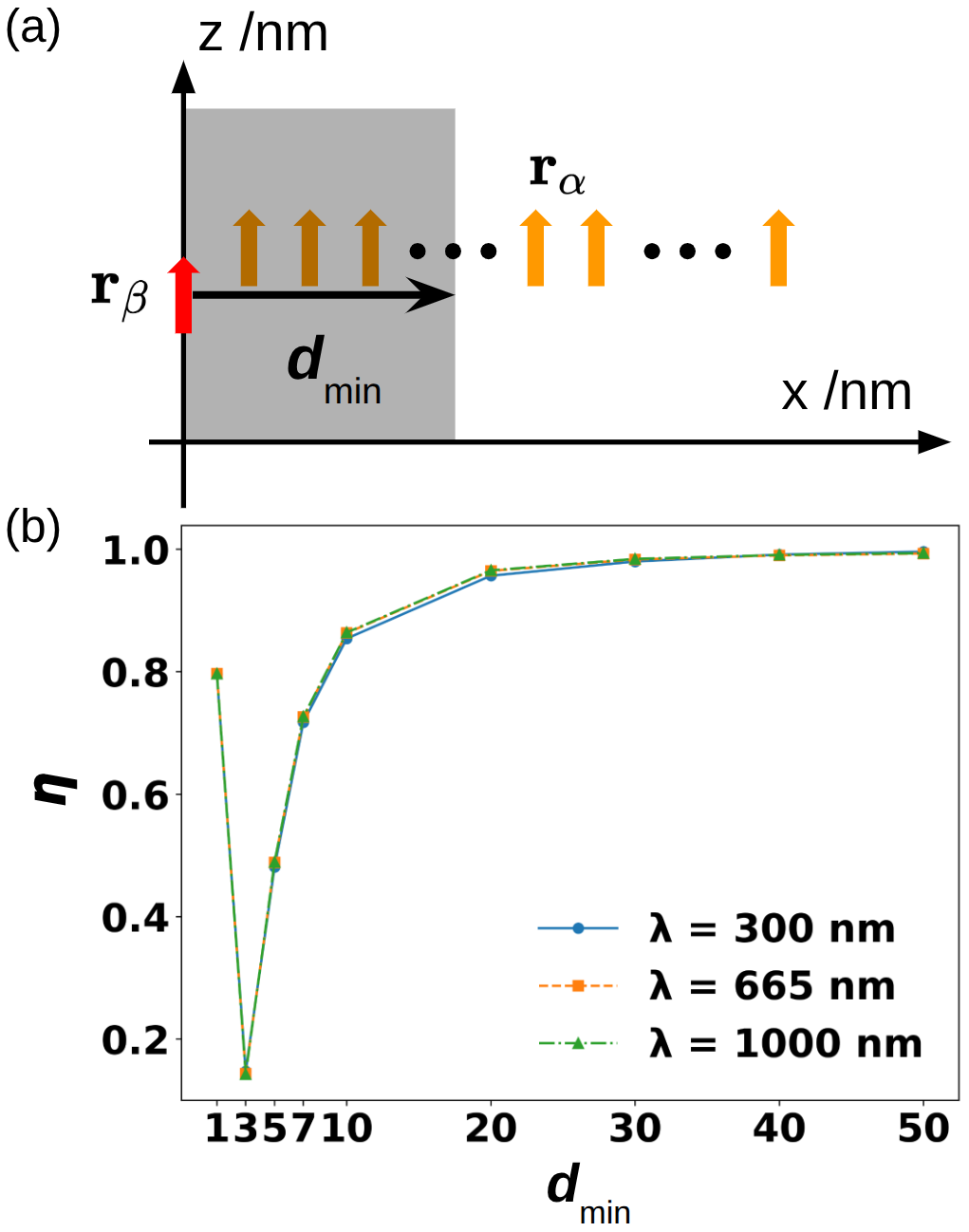}

    \caption{The calibration of dipole moment intensity in BEM. (a)Geometry used for calibration: a $z-$oriented donor dipole located at $z=2$ nm above a planar silver surface. Electric fields are sampled at a set of observation (“acceptor”) points along $x$ at $z=3$ nm. Points within a minimum separation $d$ from the donor are excluded from the least-squares fit to reduce near-source numerical artifacts in MNPBEM. (b) Reconstruction accuracy (eq~\ref{Eq:accuracy}) of the dyadic Green’s function as a function of the minimum donor–observation separation $d$ for three wavelengths ($\lambda=300, 665,\text{and}\ 1000$ nm)}
    \label{Fig:calibration}
\end{figure}

\subsection{Silver Planar Surface}

\begin{figure}[htbp]
    \centering
    \includegraphics[width=\linewidth]{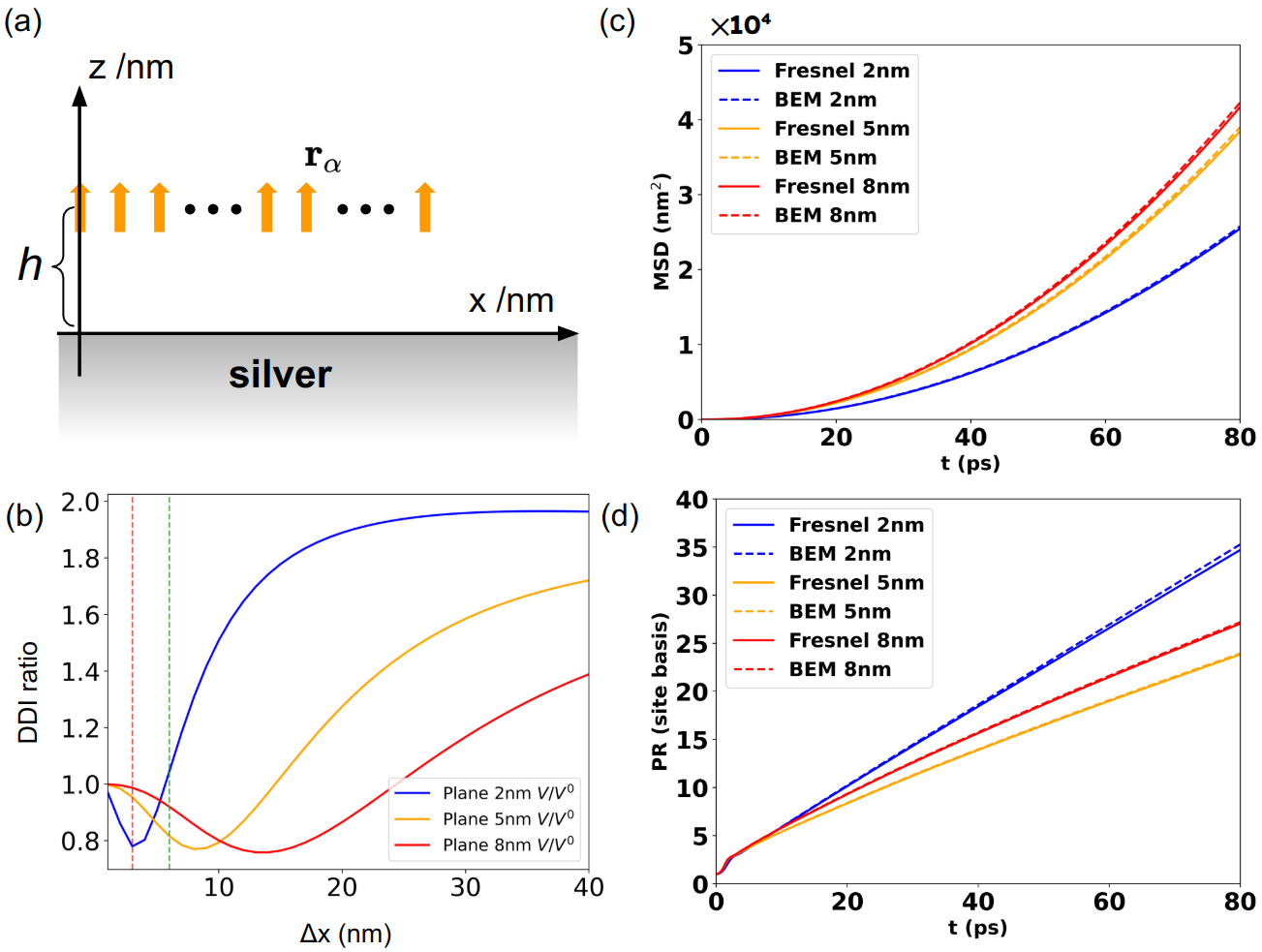}
    \caption{Exciton dynamics in molecular aggregates above silver surface with various heights. (a) Schematic illustration of molecular aggregates with 100 $z-$oriented emitters (H-aggregate) above an infinite silver planar surface.
    $h$ denotes the height of the aggregates. A single exciton is initialized at the emitter at $x=0$. $\mathbf{r}_{\alpha}$ indicates the response emitter position.
    Panel (b) shows the DDI ratio as a function of intermolecular separation ($\Delta{x}$) for different heights. The vertical red dashed line indicates the nearest-neighbor (NN) separation ($\Delta{x}=3~\text{nm}$) and the vertical green dashed line indicates the next nearest-neighbor separation ($\Delta{x}=6~\text{nm}$).
    Time evolution of the MSD (in panel c) and the PR (in panel d) as obtained by the Fresnel approach (dashed lines) and the BEM simulation (solid lines) for $h=2,5,8~\text{nm}$ (blue, orange, red). 
    The two results are identical, which further demonstrates the accuracy of the reconstructed dyadic Green's function
    through the BEM simulations. Interestingly, we observe that, while the MSD is overall reduced as $h$ decreases, the PR can be enhanced for $h=2$ nm.}
    \label{Fig:planar_silver_comparison}
\end{figure}

As our first test system, we consider a linear chain of 100 molecular emitters with an intermolecular spacing of $\Delta{x}=3~\text{nm}$ along the $x$ axis. The molecular chain is located at $h=2~\text{nm}$ above a planar silver surface (set to be the $z=0$ plane) and parallel to the surface (see Figure~\ref{Fig:planar_silver_comparison}a). 
We compare our numerical results, obtained using the calibrated BEM calculation, with the analytical solution of the dyadic Green's function evaluated via the Fresnel formulation.

\subsubsection{Calibration and Reconstruction of Dyadic Green’s Functions}

For simplicity, we implement the calibration protocol using the geometric configuration resembling a molecular chain along the $x$ axis (Figure~\ref{Fig:calibration}a). We place the electric field detectors  at $\mathbf{r}_\alpha=(x,0,3)$ (in units of nm) for $x=1,\cdots, 100$ and set an oscillating dipole source at $\mathbf{r}_\beta=(0,0,2)$ (in units of nm) oriented in the $z$ direction. Here, we offset the source dipole to avoid symmetry-related zeros in the electric field components, which provides a numerically stable dataset for subsequent calibration.
We consider three representative dipole oscillation frequencies across the UV–vis–IR wavelength range: $\lambda_\mathrm{M}=300, 665, 1000~\text{nm}$. 
Note that the calibration coefficients are calculated in the vacuum and used for the construction of the dyadic Green's function $\overline{\overline{\mathbf G}}^{\mathrm{BEM}}$ in the presence of the planar silver interface. 

To validate the dyadic Green's function, we compare
$\overline{\overline{\mathbf G}}^{\mathrm{BEM}}(\mathbf r_\alpha,\mathbf r_\beta,\omega_\mathrm{M})$
with the corresponding analytical result obtained from the Fresnel formulation (by calculating the Sommerfeld integrals), 
$\overline{\overline{\mathbf G}}^{\mathrm{Fresnel}}(\mathbf r_\alpha,\mathbf r_\beta,\omega_\mathrm{M})$.
We can quantify the accuracy in terms of the normalized inner product
\begin{equation}
\eta(\omega_\mathrm{M})
=
\frac{\sum_\alpha \left\langle
\overline{\overline{\mathbf G}}^{\mathrm{BEM}}(\mathbf r_\alpha,\mathbf r_\beta,\omega_\mathrm{M}), \,
\overline{\overline{\mathbf G}}^{\mathrm{Fresnel}}(\mathbf r_\alpha,\mathbf r_\beta,\omega_\mathrm{M})
\right\rangle_\mathrm{F}}
{\sum_\alpha \left\langle
\overline{\overline{\mathbf G}}^{\mathrm{BEM}}(\mathbf r_\alpha,\mathbf r_\beta,\omega_\mathrm{M}), \,
\overline{\overline{\mathbf G}}^{\mathrm{BEM}}(\mathbf r_\alpha,\mathbf r_\beta,\omega_\mathrm{M})
\right\rangle_\mathrm{F}},
\label{Eq:accuracy}
\end{equation}
where $\langle \mathbf A,\mathbf B\rangle_\mathrm{F} \equiv \sum_{i,j} A_{ij}^* B_{ij}$ denotes the Frobenius inner product
evaluated at the same $(\mathbf r_\alpha,\mathbf r_\beta,\omega_\mathrm{M})$ and we sum over all electric field detectors $\mathbf{r}_\alpha$.
For each $d_\mathrm{min}$, we calculate the calibration coefficient $s(\omega_\mathrm{M})$ using eq.~\ref{eq:s_closedform}, then run the BEM simulation for the electric fields and reconstruct the tensor elements of the dyadic Green's function using eq.~\ref{Eq:E_over_s}. 

Figure~\ref{Fig:calibration}b shows that the dyadic Green's function constructed by the calibrated BEM electric fields is accurate when $d_\mathrm{min}$ is sufficiently large ($d_\mathrm{min} \ge 50~\mathrm{nm}$ here).
Furthermore, for the wavelength range we consider, the accuracy does not depend on the wavelength. 
We emphasize that, while the calibration protocol uses only a subset of points ($x>d_\mathrm{min}$) to avoid the near-source oscillations, the reconstructed Green's function using the calibration coefficient shows great accuracy for all $x=[1,\cdots,100]$.


\subsubsection{Transport and Delocalization Properties}
With the validated dyadic Green's functions, we now compute the DDI strength $V_{\alpha\beta}$ and the dissipation rate $\Gamma_{\alpha\beta}$ from eqs~\ref{Eq:DDI} and \ref{Eq:Gamma}. 
Next, we propagate the quantum state vector using the effective non-Hermitian Hamiltonian to obtain the MSD and PR (eqs~\ref{Eq:MSD} and \ref{Eq:PR}). For benchmarking, we compare our results from the BEM simulation with those by the analytical solution in various configurations. 
In Figure~\ref{Fig:planar_silver_comparison}, we demonstrate that the MSD and PR as obtained by the analytical solution and the BEM simulation show almost identical results at different heights from the silver surface.

In addition to the agreement, we observe that the MSD shows a quadratic scaling with respect to time, indicating ballistic transport, and is suppressed as $h$ decreases. The PR, however, exhibits a linear scaling at long times and the slope is enhanced at $h=2~nm$. Interestingly, the PR does not show the same monotonic trend as the MSD: the PR is lowest at $h=5~\text{nm}$, but increases at $8~\text{nm}$. 

To understand this different trend between transport and delocalization, we calculate the DDI ratio, $V_{\alpha\beta}/V_{0,\alpha\beta}$, for two dipoles with an intermolecular separation $|\mathbf{r}_\alpha-\mathbf{r}_\beta|=\Delta{x}$. Here, $V_{\alpha\beta}$ and $V_{0,\alpha\beta}$ are the DDI strength (eq~\ref{Eq:DDI}) evaluated using the total dyadic Green's function $\overline{\overline{\mathbf{G}}}(\mathbf{r}_\alpha,\mathbf{r}_\beta,\omega_\mathrm{M}) $ and the vacuum dyadic Green's function $\overline{\overline{\mathbf{G}}}_0(\mathbf{r}_\alpha,\mathbf{r}_\beta,\omega_\mathrm{M})$, respectively. 
In Figure~\ref{Fig:planar_silver_comparison}, the vertical red dashed line indicates $\Delta{x}=3~\text{nm}$, which is the intermolecular distance used in our molecular chain simulation. At this separation, the DDI ratio is largest at $h=8~\text{nm}$ and smallest at $h=2~\text{nm}$, suggesting that the MSD is dominated by the nearest-neighbor coupling strength. However, at a horizontal distance of $\Delta{x}=6~\text{nm}$ (vertical green dashed line), corresponding to the next-nearest-neighbor coupling, the DDI ratio follows the descending order: $h=2~\text{nm}$,  $h=8~\text{nm}$, $h=5~\text{nm}$, which agree with the order of the PR.
These observations indicate that nearest-neighbor coupling is dominant in transport, while long-range interactions play a significant role in delocalization.
Similar trends are also observed in the nanorod system discussed in the following section.

\begin{figure}[h]
    \centering
    \includegraphics[width=1.0\linewidth]{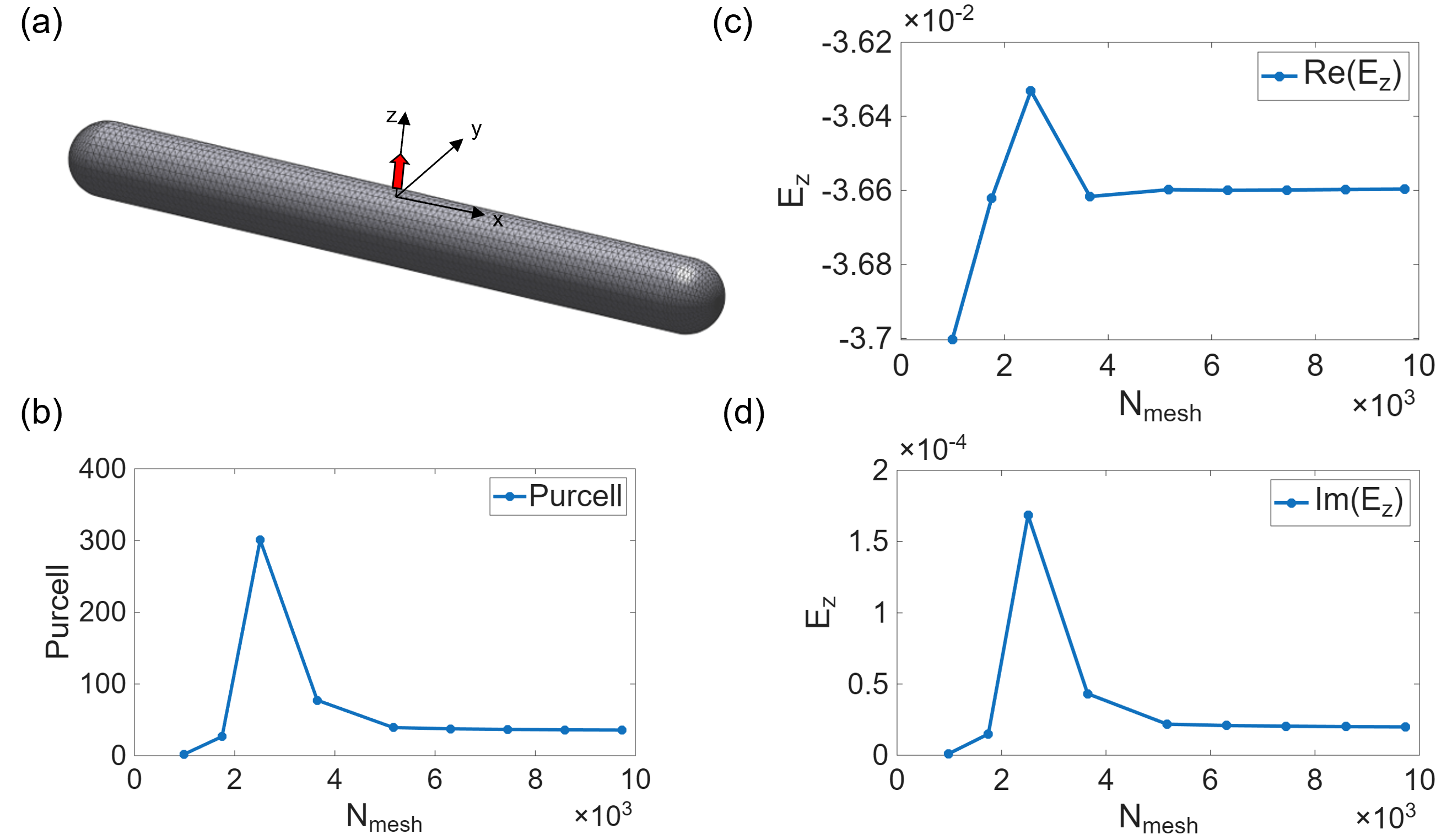}
    \caption{Convergence of the electric field and Purcell factor above a silver nanorod. 
    (a) Schematic illustration of the simulation setup where an oscillating dipole is located in the middle of the nanorod at height $h=8~\text{nm}$ and orientated in the $z$ direction. The length of the nanorod is $1000~\text{nm}$ and the radius is $10~\text{nm}$. 
    (b) The Purcell factor of the dipole above the nanorod converges with the mesh size $N_\mathrm{mesh}$ (the maximum side length of a triangular boundary element in the BEM simulation).
    At a closed observation point at $\Delta x=3~\text{nm}$, The real part (c) and imaginary part (d) of $E_z$ also converge as the mesh size increases.}
    \label{Fig:nanorod_convergence}
\end{figure}

\subsection{Silver Nanorod}
To further demonstrate the capability of the \textsc{MQED-QD} package, we now turn our attention to a test system where the analytical solution of the dyadic Green's function is not available. 
Our second test system comprises a silver nanorod (with $10~\text{nm}$ radius and $1000~\text{nm}$ length) and a molecular chain with $N_\mathrm{M}=30$ and intermolecular distance $\Delta{x}=8~\text{nm}$.
The molecular chain is placed along the longitudinal axis (the $x$ axis) and 
$h=8~\text{nm}$ above the surface of the nanorod as shown in Figure~\ref{Fig:nanorod_convergence}a. 

\begin{figure}[htbp]
    \centering
    \includegraphics[width=0.6\linewidth]{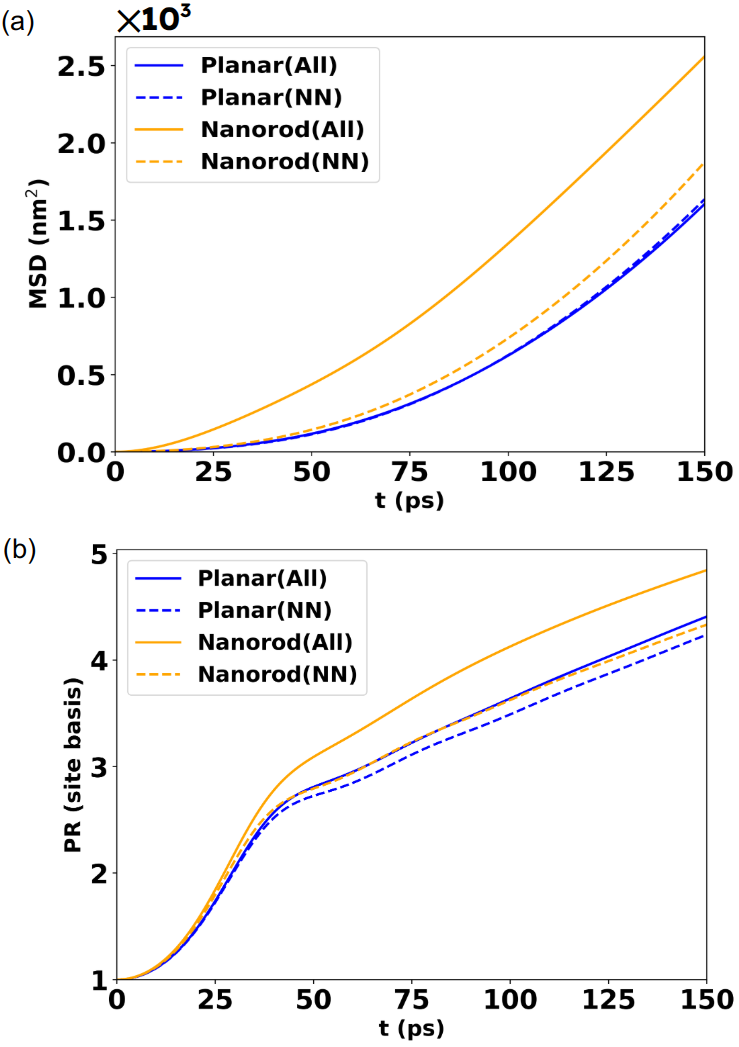}
    \caption{Quantum dynamics of $N_\mathrm{M}=30$ molecular chain above a silver nanorod (radius $10~\mathrm{nm}$, length $1000~\mathrm{nm}$) with intermolecular separation $\Delta{x}=8~\mathrm{nm}$ and height $h=8~\mathrm{nm}$. 
    (a) Mean-square displacement (MSD) as a function of time. 
    (b) Participation ratio (PR) as a function of time. 
    Solid blue: planar silver surface. Solid green: silver nanorod. 
    Dashed curves show results obtained with a nearest-neighbor (NN) truncation of the intermolecular couplings for the corresponding geometry.}
    \label{Fig:nanorod_comparison}
\end{figure}

\subsubsection{Calibration and Convergence}
We follow the calibration protocol and set the source dipole at the center of the longitudinal axis of the nanorod (see Figure~\ref{Fig:nanorod_convergence}a). 
Before determining the calibration coefficient, we need to ensure the mesh size for the BEM simulation is large enough.
Thus, we choose the closed observation point at $\Delta{x}=3~\text{nm}$ and evaluate the Purcell factor and the electric fields.
By increasing the mesh size ($N_\mathrm{mesh}$), we find the optimal mesh size of convergence for the calibration of the BEM simulation. Figure~\ref{Fig:nanorod_convergence}b--\ref{Fig:nanorod_convergence}d show that the Purcell factor and the $z$-component of the electric field converge around $N_\mathrm{mesh}\approx5000$.
Thus, we use this mesh size to determine the calibration coefficient and reconstruct the dyadic Green's function above the silver nanorod. 
Note that this optimal mesh size is sensitive to the geometry of the nanostructure and test dipole positions. 

However, unlike a planar surface, the dyadic Green's function of a nanorod does not obey translational symmetry, i.e., $\overline{\overline{\mathbf{G}}}_{ij}(\mathbf{r}_{\alpha},\mathbf{r}_{\beta},\omega)
\neq \overline{\overline{\mathbf{G}}}_{ij}(\mathbf{r}_{\alpha}-\mathbf{r}_{\beta},\omega)$. In principle, for a $N_\mathrm{M}$ molecular aggregate system, we need to perform the BEM simulations for $3(N_\mathrm{M}-1)$ source dipoles to reconstruct the two-point dyadic Green's function tensor. 
For simplicity, we choose the longitudinal length of the nanorod ($1000~\text{nm}$) to be much larger than the molecular chain, so that we can approximate the two-point dyadic Green's function using the translational symmetry. 
This assumption also relies on the setting that the molecular chain is located near the longitudinal center of the nanorod. Namely, the nanorod is effectively infinite long for the molecules. This approximation significantly reduces the computational cost for reconstructing the dyadic Green's function. 

\subsubsection{Planar Surface vs. Nanorod: Transport and Delocalization Properties}
In Figures~\ref{Fig:nanorod_comparison}a and \ref{Fig:nanorod_comparison}b, we compare the transport (MSD) and delocalization (PR) of the molecular exciton above a silver nanorod and a planar surface. 
Exciton transport above the silver nanorod is significantly enhanced compared to the planar surface, and the MSD retains a quadratic time dependence.
The PR exhibits a quadratic increase at short times, where the silver nanorod and the planar surface yield nearly identical behavior. 
At longer times, the PR transitions to a linear scaling with time, and the silver nanorod delocalizes the exciton more efficiently than the planar surface. 

To identify the origin of the transport enhancement, we performed control simulations restricted to the nearest-neighbor (NN) level ($V_{\alpha\beta}=0$ for $|\beta-\alpha| \neq 1$), thereby excluding the contribution of long-range DDI. In this truncated model (dashed lines in Figure~\ref{Fig:nanorod_comparison}), the MSD and PR dynamics for both the nanorod and the planar surface are nearly identical, indicating that the short-range DDI is equivalent for the two systems. We therefore conclude that the significantly faster exciton delocalization and enhanced transport observed with the full Hamiltonian of the silver nanorod (solid yellow lines) arise specifically from the enhanced long-range DDI.

In contrast, the planar system is insensitive to long-range DDI. A comparison between the full Hamiltonian (solid blue lines) and the truncated model (dashed blue lines in Figure~\ref{Fig:nanorod_comparison}) reveals nearly identical dynamics for both MSD and PR. This overlap indicates that, on the planar surface, exciton transport is almost entirely governed by NN interactions, with long-range DDI contributing negligibly. This result implies that effective long-range DDI is not a universal feature of all dielectric environments; rather, it requires specific geometric confinement or plasmonic guidance, highlighting the importance of tailoring nanophotonic structures to engineer long-range coupling channels.

\subsubsection{Enhancement of Long-Range DDI}

\begin{figure}[t]
    \centering
    \includegraphics[width=0.6\linewidth]{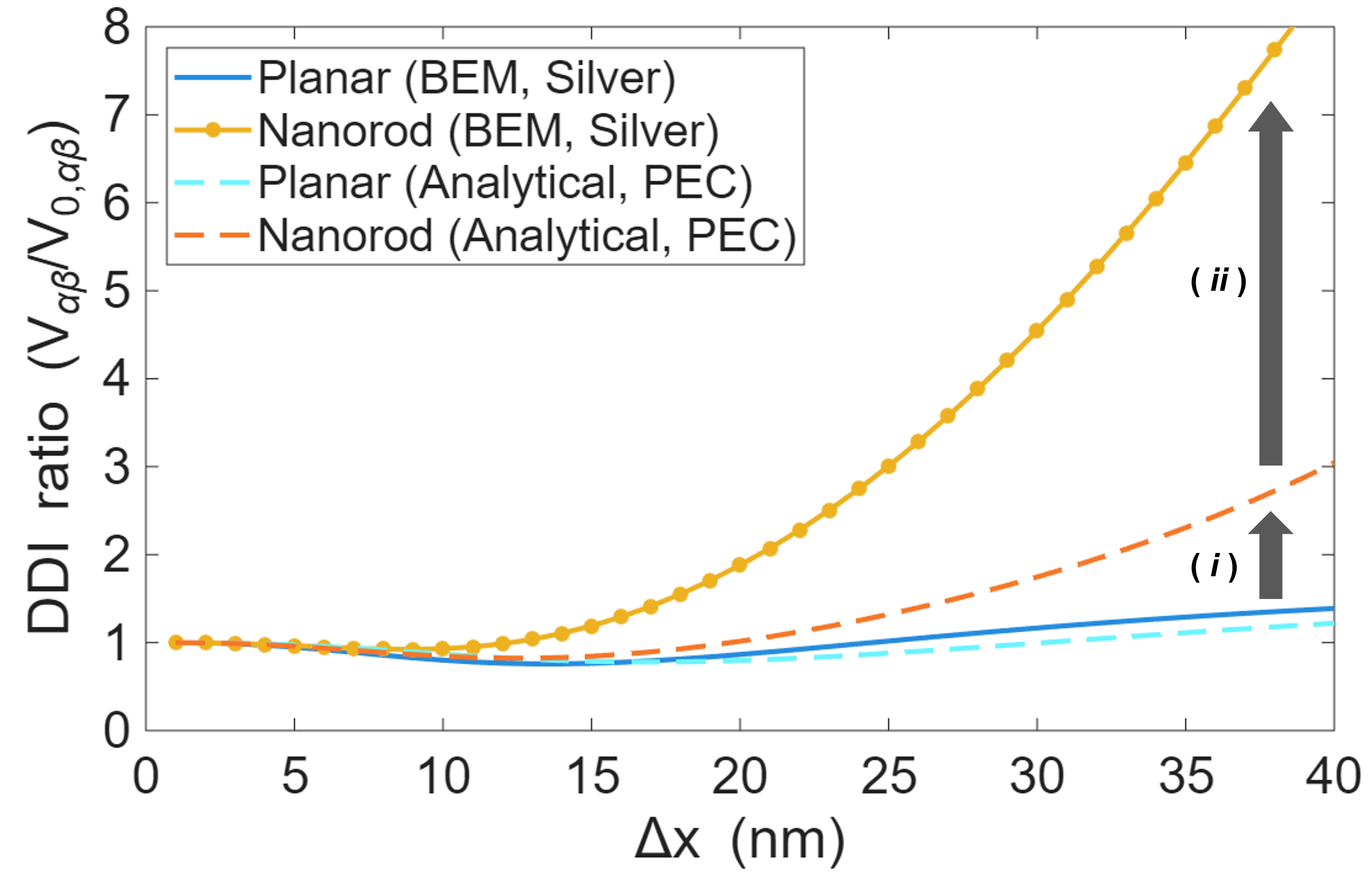}
    \caption{Enhancement of the DDI ratio driven by geometric effect and plasmonic excitation. The nearly coincident blue solid line (silver plane) and cyan dashed line (PEC plane) establish the planar baseline. Arrow (i) highlights the purely geometric enhancement achieved by transitioning from the planar geometry to a PEC nanorod (dashed orange line). Arrow (ii) demonstrates the dramatic enhancement observed upon introducing a realistic silver nanorod (solid yellow curve with markers). The enhanced long-range DDI is a two-fold combined effect driven by both SPP-mediated propagation and geometric field confinement.}
    \label{fig:enhancement_compare_V}
\end{figure}

To quantify the physical mechanisms driving the enhanced long-range interactions, we investigate the DDI ratio, $V_{\alpha\beta}/V_{0,\alpha\beta}$, for two dipoles with an intermolecular separation $|\mathbf{r}_\alpha-\mathbf{r}_\beta|=\Delta{x}$ for both the cylinder and plane systems. 
In Figure~\ref{fig:enhancement_compare_V}, we plot the DDI ratio as a function of the intermolecular separation $\Delta x $ for both the silver nanorod and silver planar surface. For the planar system, the DDI ratio remains near unity across the entire separation range shown (up to $\Delta x = 40~\text{nm}$). Since the free-space DDI scales as $(\Delta x)^{-3}$, this minimal enhancement on the planar surface indicates that long-range DDI has a negligible effect on the resulting MSD and PR dynamics, consistent with the nearly identical solid and dashed blue lines in Figure~\ref{Fig:nanorod_comparison}.

For the cylindrical system at small separations ($\Delta x < 10~\text{nm}$), the DDI ratio for the silver nanorod is nearly identical to that of the planar surface. Because the nearest-neighbor distance in our molecular chain is $8~\text{nm}$, this equivalence at short range explains why the truncated coupling models produce very similar MSD and PR dynamics for both geometries (dashed yellow and dashed blue lines in Figures~\ref{Fig:nanorod_comparison}a and \ref{Fig:nanorod_comparison}b). However, for separations exceeding $10~\text{nm}$, the DDI ratio for the silver nanorod becomes substantially larger than that of the planar system. This pronounced enhancement at larger separations directly reflects the emergence of robust long-range coupling mediated by the silver nanorod, which accounts for the enhanced MSD and PR shown by the solid yellow lines in Figure~\ref{Fig:nanorod_comparison}.


To isolate the geometric and plasmonic contributions to this long-range DDI enhancement, we introduce perfect electric conductor (PEC) models for both the plane and cylinder (see section~S2 in the Supporting Information (SI)). Notably, in Figure~\ref{fig:enhancement_compare_V}, the cyan dashed line coincides with the blue solid line, indicating that the planar silver surface behaves effectively as a PEC surface when evaluating the DDI ratio at $\hbar \omega_\mathrm{M} \approx 1.864$~eV. 

However, transitioning from the planar geometry (both silver and PEC planes) to a PEC nanorod (dashed orange line) yields a moderate increase in DDI strength, as indicated by arrow (i). 
This step isolates the purely geometric contribution arising from the radial confinement of the electric field around the cylinder (see Figure~S4 in SI). Moreover, replacing the idealized PEC boundary with a realistic silver nanorod (solid yellow curve with markers) produces a dramatic secondary enhancement at larger separations, denoted by arrow (ii). This substantial divergence from the PEC nanorod can be attributed to the plasmonic response of the material, namely the excitation and efficient longitudinal propagation of surface plasmon polariton (SPP) modes (see Figure~S5 in SI). Consequently, the robust long-range enhancement observed in the silver nanorod system is a synergistic two-fold effect, driven by SPP-mediated propagation acting in conjunction with geometric field confinement.

\section{Conclusion}\label{sec:conclusion}



In this work, we have developed an open-source package \textsc{MQED-QD} that bridges the MQED framework and open quantum system dynamics for simulating exciton transport and delocalization of molecular aggregates within complex dielectric environments. We establish a robust workflow for calculating the dyadic Green's function by numerical Maxwell solvers (specifically, BEM implemented in the MNPBEM toolbox), parametrizing the MQED Hamiltonian, and solving quantum master equations (specifically, by the QuTiP). 
We verify the accuracy of our results by comparing the quantum dynamics results with the analytical solution in an infinite planar surface geometry. 
We further demonstrate the capability of the \textsc{MQED-QD} package to investigate exciton transport near a finite silver nanorod and unravel the important role of long-range DDI to facilitate exciton delocalization and its physical origin. 

The \textsc{MQED-QD} package provides the following advantages, making it a user-friendly platform for researchers to investigate exciton dynamics in complex dielectric environments.
The package reads a Hydra-based configuration and can be executed through a small set of  terminal commands.
The QuTiP-based quantum dynamics is extensible for calculating other relevant dynamical quantities, such as coherence length and higher-order moments.
Finally, despite the lack of extensive performance benchmarking, our simulations typically complete within 30 minutes on a personal laptop, demonstrating its accessibility for high-throughput research.

Nevertheless, there are several limitations in the current version of \textsc{MQED-QD} package. First, for a nanostructure without translational symmetry, the dyadic Green's function should be reconstructed as a high-dimensional tensor indexed by all pairs of emitter indices ($\alpha, \beta$) and their three-dimensional Cartesian components ($i,j \in \{x,y,z\}$). This requirement significantly increases the computational cost compared to systems with higher symmetry. 
Second, the current implementation includes only Lindblad master equation, which assumes a weak-coupling regime between molecular aggregates and the dielectric environment and the Born-Markov approximation for the dynamical response. 
Third, the current MQED-QD package assumes a uniform molecular transition frequency. Future versions could be extended to incorporate on-site energy and DDI disorder, enabling the study of exciton transport and localization in amorphous molecular assemblies. 
Fourth, the current MQED-QD Hamiltonian is restricted to the single-excitation manifold and neglects nuclear degrees of freedom. While lifting these restrictions is formally possible, it would require deriving a more generalized MQED-based effective Hamiltonian to capture multi-exciton dynamics and vibronic couplings, which would inherently lead to more complex quantum dynamics.

We look forward to expanding the functionalities of the \textsc{MQED-QD} package to address a broader range of complex physical phenomena. Specifically, we aim to develop the following features. 
(\emph{i}) The integration of wavefunction-based methods\cite{chuang_2024} or few-mode models\cite{sanchez2022few,chuang_2024_2} is essential for capturing non-Markovian quantum dynamics, which is critical for accurately simulating strongly coupled plasmon--polariton systems.
(\emph{ii}) Incorporating nuclear degrees of freedom via the hierarchical equations of motion (HEOM) or tensor-network-based methods will enable the investigation of vibronic coupling and non-adiabatic effects at the light--matter interface.
(\emph{iii}) Developing more efficient machine-learning-based surrogate models will facilitate the bypass of direct dyadic Green's function constructions. 
We expect these developments will bridge the gap between the nanophotonics and quantum dynamics communities.




\begin{acknowledgement}
Liu, Wang and Chen thank the support provided by the University of Notre Dame and the Asia Research Collaboration grant of Notre Dame Global. The authors also thank Qian-Rui Huang and Liang-Yan Hsu at the Institute of Atomic and Molecular Sciences, Academia Sinica, for helpful discussions and comments.

\end{acknowledgement}

\begin{suppinfo}
Example of YAML File; Analysis of the Dipole-Dipole Interaction Strength.

The source code of the MQED-QD package is openly available on our Github repository: \texttt{https://github.com/MQED-QD/Macroscopic-Quantum-Electrodynamics}

\end{suppinfo}

\bibliography{achemso-demo}

\end{document}